%% file: majun.tex
\documentclass{emulateapj}

\received{} \accepted{}

\shorttitle{Globular Clusters in M81} \shortauthors{Ma et al.}

\begin{document}

\title{Spectral Energy Distributions of M81 Globular Clusters in BATC
Multicolor Survey}

\author{
Jun Ma\altaffilmark{1}, Xu Zhou\altaffilmark{1}, David
Burstein\altaffilmark{2}, Jiansheng Chen\altaffilmark{1}, Zhaoji
Jiang\altaffilmark{1}, Zhenyu Wu\altaffilmark{1},Jianghua
Wu\altaffilmark{1}}

\altaffiltext{1}{National Astronomical Observatories, Chinese
Academy of Sciences, Beijing, 100012, P. R. China;
majun@vega.bac.pku.edu.cn}

\altaffiltext{2}{Department of Physics and Astronomy, Box 871504,
Arizona State University, Tempe, AZ  85287--1504}


\begin{abstract}

In this paper, we give the spectral energy distributions of 42 M81
globular clusters in 13 intermediate-band filters from 4000 to
10000{\AA}, using the CCD images of M81 observed as part of the
BATC multicolor survey of the Sky. The BATC multicolor filter
system is specifically designed to exclude most of the bright and
variable night-sky emission lines including the OH forest. Hence, it
can present accurate SEDs of the observed objects. These
spectral energy distributions are low-resolution spectra, and can
reflect the stellar populations of the globular clusters. This
paper confirms the conclusions of \citet{sbkhp02} that, M81
contains clusters as young as a few Gyrs, which also were observed
in both M31 and M33.
\end{abstract}


\section{Introduction}

The study of globular clusters (GCs) plays an important role in
our understanding of the evolution and history of galaxies. They
are bright and easily identifiable star clusters typically with
homogeneous abundances and ages. The Galactic GCs, the stars of
which are thought to be among the oldest stars in the universe
provide important information regarding the minimum age of the
universe and the early formation history of our Galaxy. However,
we also find that the GC system of our neighboring galaxy, M31,
contains at least 20 young GCs ranging in age from 100 Myr to
$\sim 5$ Gyr \citep{bur04,b05}.

Except for the Local Group galaxies, M81 is one of the nearest
large spirals outside the Local Group. As such, its globular
cluster system has come under recent detailed scrutiny.
\citet{pr95} first attempted to identify GCs in M81 from
ground-based images, sifting through over 3700 objects in a
$50\arcmin$ diameter field centered on M81. They found $~70$ GC
candidates within 11 kpc galactocentric radius. \citet{pbh95} then
confirmed 25 as M81 GCs on basis of spectroscopy of 82 bright GC
candidates in the M81 field. \citet{sbkhp02} obtained
moderate-resolution spectroscopy for 16 of the Perelmuter \&
Racine M81 GC candidates, and found that all of these are GCs.

Recently, \citet{cft01} discovered 114 compact star clusters in
M81 from $B$-, $V$-, and $I$-band {\it {Hubble Space Telescope}}
(HST) Wide Field Planetary Camera 2 images in eight fields,
covering a total area of 40 arcmin, 54 of which are new GCs. Using
these 95 M81 GCs, \citet{ma05} presented that the intrinsic $B$
and $V$ colors and metallicities of these GCs are bimodal, with
metallicity peaks at $\rm {[Fe/H]}\approx -1.45$ and $-0.53$,
similar to what we find for the Milky Way and M31 GCs. In this
paper we present new spectral energy distributions (SEDs) for 42
of these GCs, using M81 images observed as part by galaxy
calibration program of the Beijing-Arizona-Taiwan-Connecticut
(BATC) multicolor sky survey \citep[e.g.,][]{fan96,zheng99}. The
BATC filters are custom-designed set of 15 intermediate-band
filters to do spectrophotometry for preselected 1 deg$^{2}$
regions of the northern sky.

Details of our observations and data reduction are given in \S~2.
\S~3 gives our summary.

\section {Observations and Data Reduction}

\subsection {The Sample of GCs}

\citet{ma05} studied the intrinsic $B$ and $V$ colors and
metallicities of 95 M81 GCs. In order to study the stellar
populations of these GCs, we extracted 311 images of M81 field as
part of the BATC multicolor survey of the sky, taken in 13
intermediate-band filters with a total exposure time of $\sim 100$
hours from February 5, 1995 to April 30, 2002. Multiple images of
the same filter were combined to improve the signal-to-noise
ratio. While the SEDs of the sample GCs brighter than $V\sim 20$
mag can be obtained in the BATC multicolor system, we are
constrained in obtaining full SEDs for these GCs by the limited
field of view around M81, as well as by some of these GCs being in
the high background of M81 itself. As such, we have obtained SEDs
in 13 BATC filters for 42 of the 95 GCs previously presented.

\subsection{Observations and data reduction}

The BATC multicolor survey uses a Ford Aerospace $2048\times 2048$
CCD camera with 15 $\mu$m pixel size on the 0.6/0.9m f/3 Schmidt
telescope of the Xinglong Station of the National Astronomical
Observatories, giving a CCD field of view of $58^{\prime}$ $\times
$ $58^{\prime}$ with a pixel size of
$1\arcsec{\mbox{}\hspace{-0.15cm}.} 7$. The typical seeing of the
Xinglong station is $2\arcsec$. The BATC multicolor filter system,
which was specifically designed to avoid contamination from the
brightest and most variable night sky emission lines, includes 15
intermediate-band filters covering 3300{\AA} to 1$\mu$.
Calibrations of these images are made using observations of four
$F$ sub-dwarfs, HD~19445, HD~84937, BD~${+26^{\circ}2606}$, and
BD~${+17^{\circ}4708}$, all taken from \citet{ok83}.  Hence, our
magnitudes are defined in a way similar to the spectrophotometric
AB magnitude system that is the Oke \& Gunn $\tilde{f_{\nu}}$
monochromatic system. BATC magnitudes are defined on the AB
magnitude system as

\begin{equation}
m_{\rm batc}=-2.5{\rm log}\tilde{F_{\nu}}-48.60,
\end{equation}

\noindent where $\tilde{F_{\nu}}$ is the appropriately averaged
monochromatic flux in unit of erg s$^{-1}$ cm$^{-2}$ Hz$^{-1}$ at
the effective wavelength of the specific passband. In the BATC
system \citep{yan00}, $\tilde{F_{\nu}}$ is defined as

\begin{equation}
\tilde{F_{\nu}}=\frac{\int{d} ({\rm log}\nu)f_{\nu}r_{\nu}}
{\int{d} ({\rm log}\nu)r_{\nu}},
\end{equation}

\noindent which links the magnitude to the number of photons
detected by the CCD rather than to the input flux \citep{fuku96}.
In equation (2), $r_{\nu}$ is the system's response, $f_{\nu}$ is
the SEDs of the source.

Of the 15 BATC filters, we did not use the two bluest filters.
Data reduction of the CCD data proceeds with removal of bias
subtraction and flat-fielding with dome flats. These steps were
performed with our custom-made, automatic data reduction software,
PIPELINE I, developed for the BATC multicolor sky survey
\citep{fan96,zheng99}. The dome flat-field images were taken by
using a diffuser plate in front of the correcting plate of the
Schmidt telescope, a flatfielding technique which has been
verified with the photometry we have done on other galaxies and
fields of view
\citep[e.g.,][]{fan96,zheng99,wu02,yan00,zhou01,zhou04}.
Spectrophotometric calibration of the M81 images using the
Oke-Gunn standard stars is done during photometric nights
\citep[see details from][]{yan00, zhou01}.

Using the images of the standard stars observed on photometric
nights, we derived iteratively the atmospheric extinction curves
and the variation of these extinction coefficients with time
\citep[cf.][]{yan00,zhou01}. The extinction coefficients at any
given time in a night $[K+ \Delta K (UT)]$ and the zero points of
the instrumental magnitudes ($C$) are obtained by

\begin{equation}
m_{\rm batc}=m_{\rm inst}+[K+\Delta K(UT)]X+C,
\end{equation}

\noindent where $X$ is the air mass. The instrumental magnitudes
($m_{\rm inst}$) of the selected bright, isolated and unsaturated
stars on the M81 field images of the same photometric nights can
be readily transformed to the BATC AB magnitude system ($m_{\rm
batc}$). The calibrated magnitudes of these stars are obtained on
the photometric nights, which are then used as secondary standards
to uniformly combine images from calibrated nights to those taken
during non-photometric weather. Table~1 lists the parameters of
the BATC multicolor filter system and the statistics of
observations. Column 6 of Table~1 gives the scatter, in
magnitudes, for the photometric observations of the four primary
standard stars in each filter.

\subsection{Integrated photometry}

For each M81 GC, the PHOT routine in DAOPHOT \citep{stet87} is
used to obtain magnitudes. To avoid contamination from nearby
objects, we adopt a small aperture of
$6\arcsec{\mbox{}\hspace{-0.15cm}.}8$ corresponding to a diameter
of 4 pixels in the Ford CCD. Aperture corrections are determined
as follows, using the isolated M81 GC, Is40165: determining the
magnitude differences between photometric diameters 4 and 10
pixels in each of the 13 BATC filters. Inner and outer radii of
the sky apertures are from 4 to 7 pixels for a diameter of 4
pixels, and from 6 to 10 pixels for a diameter of 10 pixels. The
SEDs obtained in this manner are given in Table~2. Column 1 is
GC's name taken from \citet{pbh95}, \citet{sbkhp02} and
\citet{cft01}. Column 2 to column 14 give the magnitudes of the 13
BATC passbands observed. The second line for each GC gives the
$1-\sigma$ errors in magnitudes for the corresponding passband.
The errors for each filter are given by DAOPHOT. Magnitudes in the
BATC03 filter could not be obtained for Is50286, Id50357, Id70319,
SBKHP16 and  CFT41 owing to low signal-to-noise ratios. Because of
low angular resolution, given the Schmidt pixel size of
$1\arcsec{\mbox{}\hspace{-0.15cm}.} 7$, the different sizes of
different clusters are not evident in our CCD images.

\subsection{Comparison with previous photometry}

\citet{zhou03} presented the relationships between the BATC
intermediate-band system and $UBVRI$ broadband system using the
standard stars catalogs of \citet{land83,land92} and
\citet{gtj00}. The coefficients of two relationships are showed by
equations (4) and (5) below:

\begin{equation}
m_B=m_{04}+0.2201(m_{03}-m_{05})+0.1278\pm0.076,
\end{equation}

\begin{equation}
m_V=m_{07}+0.3292(m_{06}-m_{08})+0.0476\pm0.027.
\end{equation}

Using equations (4) and (5), we transformed the magnitudes of 42
GCs in BATC03, BATC04 and BATC05 bands to ones in the $B$ band,
and in BATC06, BATC07 and BATC08 bands to ones in $V$ band. Figure
1 plots the comparison of $V$ (BATC) and ($B$$-$$V$) (BATC)
photometry with previously published measurements of \citet{pr95}
and \citet{cft01}. In this figure, our  magnitudes/colors are on
the x-axis, the difference between our and \citet{pr95} and
\citet{cft01} magnitudes/colors are on the y-axis. Table 3 lists
this comparison. The mean $V$ magnitude and color differences (in
the sense of this paper $-$ \citet{pr95} and \citet{cft01}) are
$<\Delta V> =-0.116\pm0.028$ and $<\Delta (B-V)>=-0.017\pm 0.027$,
respectively. The uncertainties in $<\Delta V>$ and $<\Delta
(B-V)>$ are calculated by

\begin{equation}
\sqrt{\frac{\Sigma(<\Delta V>-\overline{<\Delta V>})^2}{N(N-1)}},
\end{equation}
and
\begin{equation}
\sqrt{\frac{\Sigma(<\Delta (B-V)>-\overline{<\Delta
(B-V)>})^2}{N(N-1)}}.
\end{equation}

Uncertainties in $B$ (BATC) and $V$ (BATC) have
been added linearly, i.e.
$\sigma_B=\sigma_{04}+0.2201(\sigma_{03}+\sigma_{05})$, and
$\sigma_V=\sigma_{07}+0.3292(\sigma_{06}+\sigma_{08})$, to reflect
the errors in the three filter measurements. For the colors, we
add the errors in quadrature, i.e.
$\sigma_{(B-V)}={(\sigma_B2+\sigma_V2)}^{1/2}$. From Figure 1
and Table 3, it can be seen that there is good agreement in the
photometric measurements.

\begin{figure*}
\begin{center}
\centerline{\includegraphics[angle=-90,width=120mm]{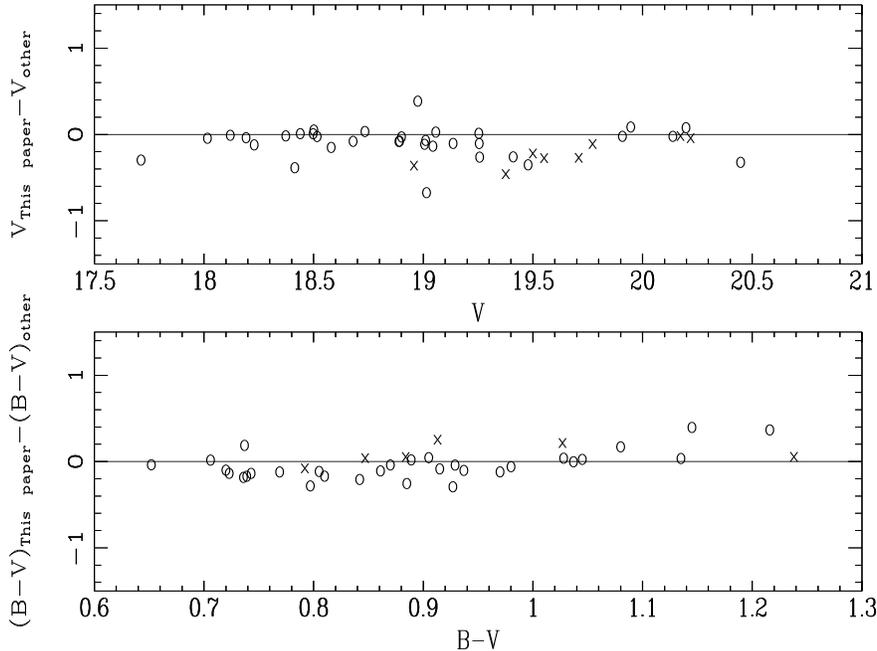}}
\caption{Comparison of cluster photometry with previous
measurements by \citet{pr95}, shown as open circles, and
\citet{cft01}, shown as crosses.} \label{fig1}
\end{center}
\end{figure*}

\subsection{Reddening}

In order to obtain intrinsic SEDs for the sample GCs, the
photometric data are corrected for the reddening from the
foreground extinction contribution of the Milky Way and for the
internal reddening due to varying optical paths through the disk
of M81. The total reddening determination for the M81 field (the
foreground plus M81 contribution) has been measured by a number of
authors \citep[e.g.,][]{fwm94,kong00}. We only mention here that,
\citet{kong00} obtained the reddening maps of M81 field based on
the images observed by the BATC multicolor sky survey in the 13
intermediate-band filters from 3800 to 10000 \AA. To determine the
metallicity, age, and reddening distributions for the M81 field,
\citet{kong00} found the best match between the observed colors
and the predictions from single stellar population models of
\citet{bc96}. A map of the interstellar reddening in a substantial
portion of M81 was obtained. For a few clusters that fall near the
edges of the images, \citet{kong00} did not obtain reddenings. For
these clusters we adopt a mean reddening value of 0.13 as
\citet{cft01} did.  The local reddening values for these GCs are
listed in column (4) of Table 1 in \citet{ma05}. Figure 2 plots
the intrinsic SEDs of 42 GCs (relative to the flux of filter
BATC08) in the 13 BATC intermediate-band filters.

\begin{figure*}
\begin{center}
\centerline{\includegraphics[angle=-90,width=170mm]{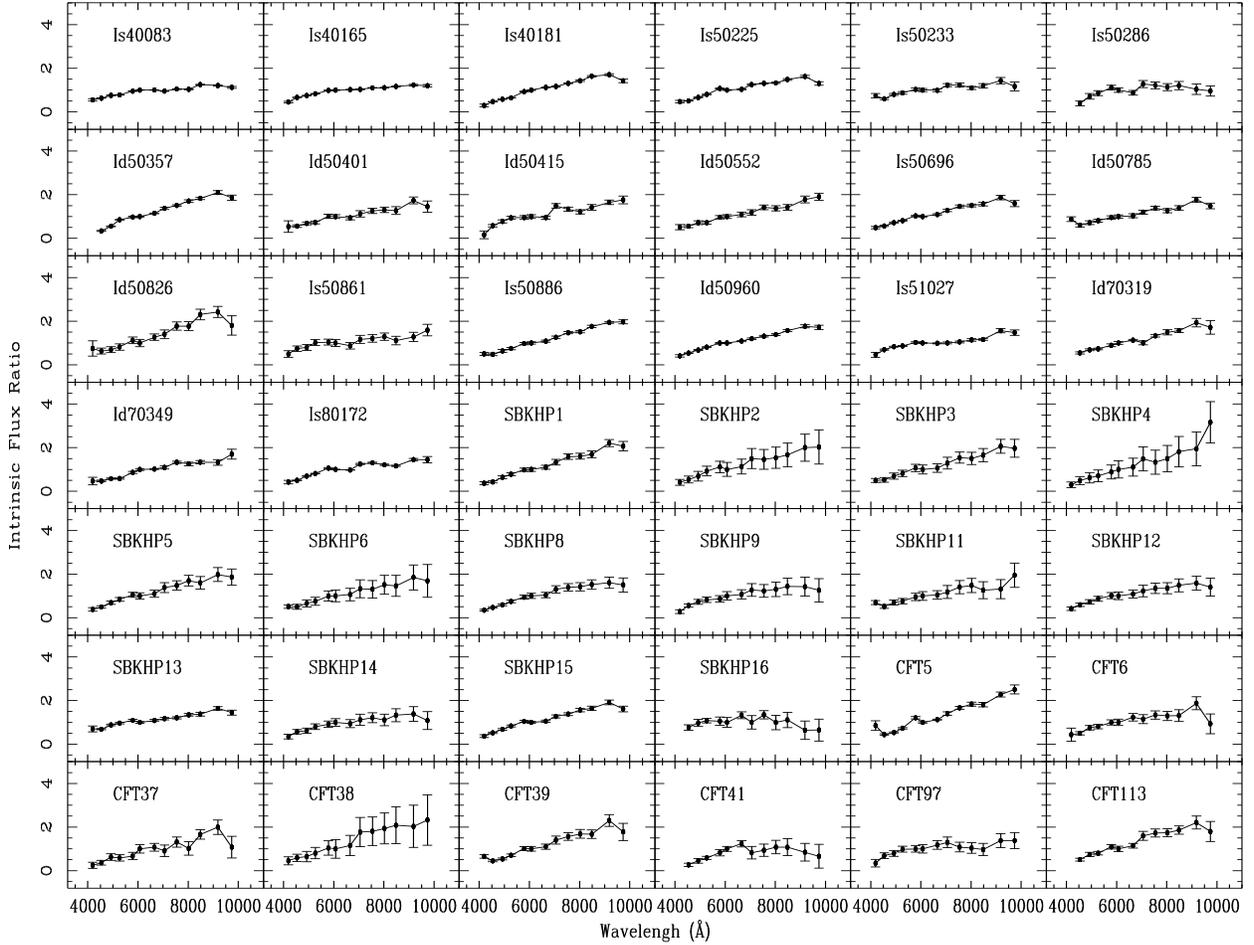}}
\caption{Intrinsic spectrophotometric energy distributions for 42
GCs in M81.} \label{fig2}
\end{center}
\end{figure*}


\subsection{Analysis of the SEDs of the GCs}

\citet{sbkhp02} observed moderate-resolution spectroscopy of 16
M81 GCs using the Low Rosolution Imaging Spectrograph on the Keck
I telescope. By comparing between the observed age-sensitive index
$\rm H\beta$ against Mg2 and isochrones from the evolutionary
synthesis models from \citet{worthey94} and from \citet{fab95},
\citet{sbkhp02} find that SBKHP15\footnote{In fact, SBKHP15 should
be SBKHP16. We referred the Table 6 of \citet{pr95} and found that
the R.A. (J2000) and Decl. (J2000) of ID 50867 are 09:55:40.194
and 69:07:30.82, and the R.A. (J2000) and Decl. (J2000) of ID
50889 are 09:55:51.995 and 69:07:39.32, i.e. the R.A. (J2000) and
Decl. (J2000) of SBKHP15 and 16 should be exchanged in
\citet{sbkhp02}.} is younger than the other GCs. As we know that,
lower metallicity and younger age can make the fluxes in longer
filter bands to be lower. The metallicity of this cluster is
nearly the same as one of SBKHP8. So, from the SEDs of Figure 2,
we can conclude that SBKHP16 is younger than SBKHP8, as its SED
is lower than those of SBKHP8 in longer filter bands. In particular,
SBKHP16 has very low fluxes in the BATC14 and BATC15 filter bands. We
compare its SEDs with ones of SBKHP8, and find that, the intrinsic
flux (relative to the flux of BATC08 filter band) is 1.607 versus
0.641 in BATC14 filter band, and 1.505 versus 0.649 in BATC15
filter band, nearly 2.5 times.

\subsection{Ages}

A single GC is a stellar population having a single age and
chemical abundance. Globular clusters are ideal systems to be
characterized by simple stellar populations (SSPs) models.

BC96 models \citep{bc96} are given for simple stellar populations
(SSPs) of metallicities $Z=0.0004, 0.004, 0.008, 0.02, 0.05$, and
$0.1$. These models are based on the Padova group evolutionary
tracks \citep{bres93,fbbc94,gbcbn96}, which use the radiative
opacities of \citet{irw92} together with a helium abundance
$Y=2.5Z+0.23$ (The reference solar metallicity is $Z_\odot=0.02$).
BC96 models further use the \citet{lcb97} standard star library.
The ages in the BC96 models range from 0 to 20 Gyr. A
\citet{salp55} IMF of $\Phi(M)=A \times M^{-\alpha}$ with
$\alpha=2.35$ is used with a normalization constant $A=1$, a lower
cutoff mass $M_{\rm l}=0.1M_{\odot}$ and an upper cutoff mass
$M_{\rm u}=125M_{\odot}$.

To proceed with the comparisons, we first convolve the SEDs of
BC96 models with the BATC filter profiles to obtain the optical
and near-infrared integrated luminosities. The integrated
luminosities $L_{\lambda_i}(t,Z)$ of the $i$th BATC filter can be
calculated as

\begin{equation}
L_{\lambda_i}(t,Z) =\frac{\int
F_{\lambda}(t,Z)\varphi_i(\lambda)d\lambda} {\int
\varphi_i(\lambda)d\lambda},
\end{equation}

\noindent where $F_{\lambda}(t,Z)$ is the SED at age $t$ in
metallicity $Z$ model, $\varphi_i(\lambda)$ is the response
functions of the $i$th filter of the BATC filter system ($i=3, 4,
\cdot\cdot\cdot, 15$), respectively. All integrated colors of BC96
models are calculated relative to the BATC08 filter band
($\lambda=6075${\AA}):

\begin{equation}
\label{color}
C_{\lambda_i}(t,Z)={L_{\lambda_i}(t,Z)}/{L_{6075}(t,Z)}.
\end{equation}

\noindent From this equation, we can obtain model
intermediate-band colors for SSP models of different
metallicities.

In order to study the stellar populations of these GCs, we use 5
ages of 1, 2, 3, 8 and 16 Gyrs of BC96 SSP models of two
metallicities: a metal-poor model of $Z=0.0004$, and a more
metal-rich model of $Z=0.004$.  The best SED-fit between a
globular cluster and SSP models is found by minimizing the color
differences between intrinsic integrated color of a cluster and
integrated color of models:

\begin{equation}
R^2(n,t,Z)=\frac{\sum_{i=3}^{15}{[C_{\lambda_i}^{\rm
intr}(n)-C_{\lambda_i}^ {\rm ssp}(t, Z)]^2}/\sigma_{i}^{2}}
{{\sum_{i=3}^{15}}{1/\sigma_{i}^{2}}},
\end{equation}

\noindent where $C_{\lambda_i}^{\rm ssp}(t, Z)$ represents the
integrated color in the $i$th filter of a SSP at age $t$ in a
metallicity $Z$ model. $C_{\lambda_i}^{\rm intr}(n)$ is the
intrinsic integrated color for a cluster. The differences are
weighted by $1/{\sigma_i}^{2}$, where the ${\sigma_i}$'s are
observational uncertainties of the passbands. Figure 3 shows the
results of SED-fits, in which filled circle represents the
intrinsic integrated color of a cluster, and the thick line
represents the best fit of the integrated color of a SSP model.

\begin{figure*}
\begin{center}
\centerline{\includegraphics[angle=-90,width=170mm]{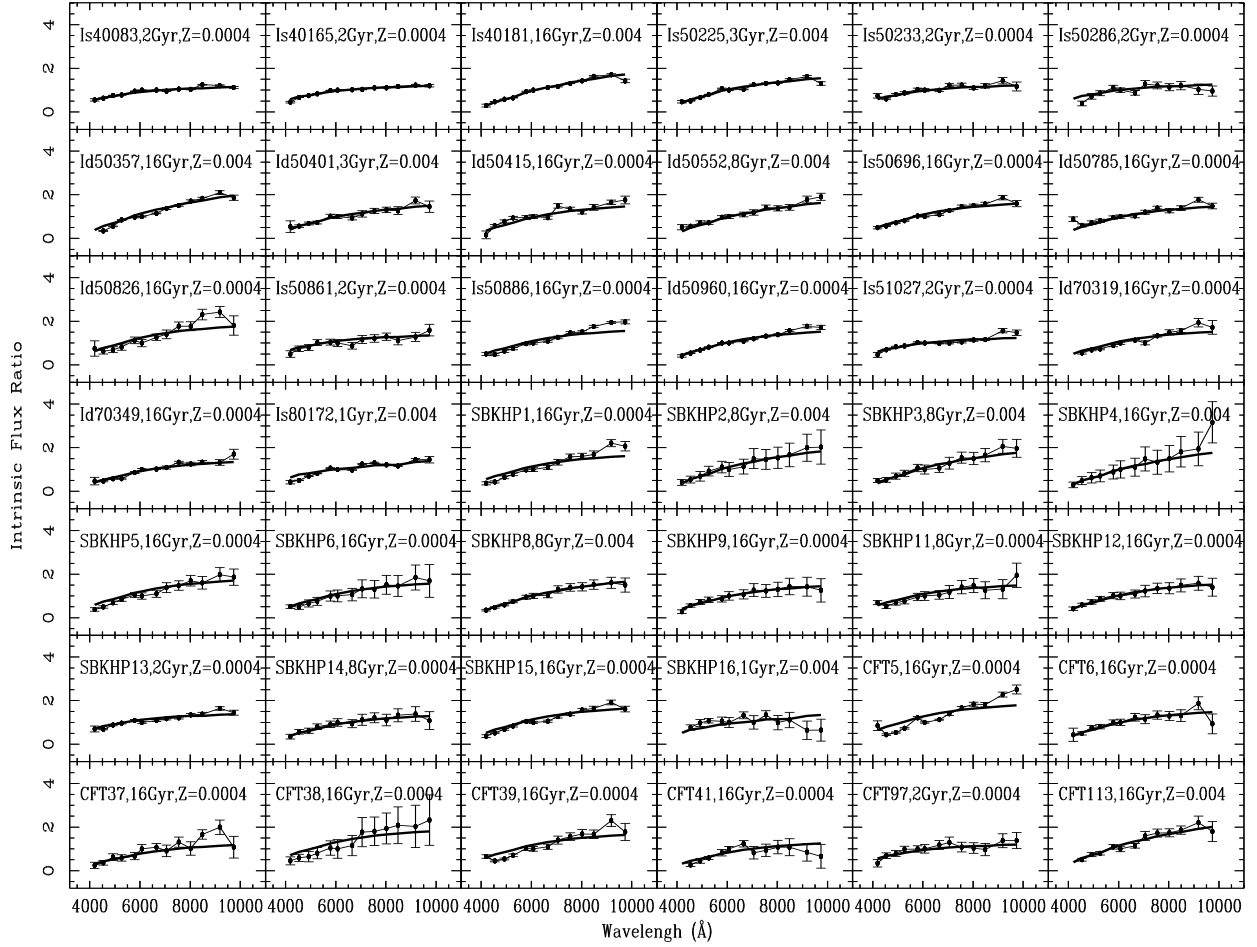}}
\caption{Map of the fit of the integrated color of a SSP model
with intrinsic integrated color for sample GCs. Filled circle
represents the intrinsic integrated color of a GC, and the thick
line represents the best fit of the integrated color of a SSP
model.} \label{fig3}
\end{center}
\end{figure*}

From Figure 3, we can see that, of these 42 M81 GCs, there are 11
for which our estimates give ages younger than 8 Gyrs. The results
tell us that, M81 includes a population of intermediate-age GCs
with ages of a few Gyr. Similar clusters have been observed in
both M31 and M33
\citep{brodie90,brodie91,jiang03,bb04,bur04,ppb05,
sarajedini98,sarajedini00,ma02}. For SBKHP16, \citet{sbkhp02}
derived its age is between 1.5 and 3 Gyrs, and our result is
consistent with this estimte, given an age of between 1 and 2
Gyrs. Our results also shows that the age of SBKHP13 is between 2
and 3 Gyrs, which was not presented by \citet{sbkhp02}, because
the value of index $\rm H\beta$ was not derived in
\citet{sbkhp02}. The ages of the other GCs of \citet{sbkhp02}
obtained in this paper are also fully consistent with the results
of \citet{sbkhp02}.

\section{Summary}

We have obtained SEDs of 42 M81 GCs in 13 intermediate-band
filters with the BATC 0.6/0.9m Schmidt telescope. The BATC filter
system is specifically designed to exclude most of the bright and
variable night-sky emission lines including the OH forest, and it
can present the accurate SEDs of the observed objects. This paper
confirms the conclusions of \citet{sbkhp02} that, M81 contains
clusters as young as a few Gyrs. Such young GCs have also been
observed in both M31 and M33
\citep{brodie90,brodie91,jiang03,bb04,bur04,ppb05,
sarajedini98,sarajedini00,ma02}.

\acknowledgments We would like to thank the anonymous referee for
his/her insightful comments and suggestions that improved this
paper very much. This work has been supported by the Chinese
National Science Foundation, No. 10473012 and by the Chinese
National Key Basic Research Science Foundation (NKBRSF
TG199075402).

\clearpage
\input{table1.tex}
\clearpage
\input{table2a.tex}
\clearpage
\input{table2b.tex}
\clearpage
\input{table3.tex}
\end{document}

%% file: table1.tex
\setcounter{table}{0}
\begin{table}
\begin{center}
\caption{Parameters of the BATC Filters and Statistics of
Observations for M81 Field}
\label{tab correlation}
\begin{tabular}{cccccc}
\tableline
\tableline
No. & Name & cw(\AA)$^a$ & Exp.(hr) & N.img$^b$ & rms$^c$\\
\tableline
1  & BATC03 & 4210   & 03:53 & 14 & 0.004\\
2  & BATC04 & 4546   & 12:20 & 39 & 0.013\\
3  & BATC05 & 4872   & 06:10 & 21 & 0.005\\
4  & BATC06 & 5250   & 06:05 & 19 & 0.005\\
5  & BATC07 & 5785   & 05:12 & 18 & 0.004\\
6  & BATC08 & 6075   & 04:00 & 12 & 0.006\\
7  & BATC09 & 6710   & 06:00 & 18 & 0.006\\
8  & BATC10 & 7010   & 05:20 & 16 & 0.007\\
9  & BATC11 & 7530   & 05:40 & 17 & 0.013\\
10 & BATC12 & 8000   & 05:20 & 16 & 0.006\\
11 & BATC13 & 8510   & 15:00 & 45 & 0.005\\
12 & BATC14 & 9170   & 16:40 & 50 & 0.036\\
13 & BATC15 & 9720   & 08:40 & 26 & 0.039\\
\tableline
\end{tabular}\\
\end{center}
{$^a$ Central wavelength for each BATC filter; $^b$ Image numbers
for each BATC filter; $^c$ Calibration error, in magnitude, for
each filter as obtained from the standard stars.}
\end{table}

%% file: table2a.tex
\setcounter{table}{1}
\begin{table}
\begin{center}
\caption{SEDs of 42 Globular Clusters in M81} \label{tab
correlation}
\begin{tabular}{cccccccccccccc}
\tableline \tableline
$\rm {Cluster}$  & 03  &  04 &  05 &  06 &  07 &  08 &  09 &  10 &  11 &  12 &  13 &  14 &  15\\
(1)    & (2) & (3) & (4) & (5) & (6) & (7) & (8) & (9) & (10) & (11) & (12) & (13) & (14)\\
\tableline
 Is40083 &  18.99 &  18.80 &  18.57 &  18.49 &  18.21 &  18.13 &  18.10 &  18.13 &  18.00 &  18.00 &  17.77 &  17.78 &  17.83\\
 \nodata &  0.100 &  0.013 &  0.013 &  0.011 &  0.015 &  0.008 &  0.007 &  0.020 &  0.015 &  0.018 &  0.017 &  0.029 &  0.041\\
 Is40165 &  19.09 &  18.63 &  18.46 &  18.30 &  18.05 &  18.02 &  17.97 &  17.93 &  17.83 &  17.81 &  17.73 &  17.64 &  17.65\\
 \nodata &  0.095 &  0.012 &  0.011 &  0.012 &  0.011 &  0.013 &  0.007 &  0.015 &  0.012 &  0.016 &  0.019 &  0.031 &  0.047\\
 Is40181 &  20.11 &  19.56 &  19.29 &  19.13 &  18.67 &  18.57 &  18.42 &  18.35 &  18.20 &  18.08 &  17.92 &  17.84 &  18.02\\
      &  0.245 &  0.024 &  0.016 &  0.018 &  0.020 &  0.012 &  0.008 &  0.024 &  0.014 &  0.022 &  0.019 &  0.028 &  0.062\\
 Is50225 &  19.23 &  19.16 &  18.85 &  18.63 &  18.31 &  18.37 &  18.33 &  18.12 &  18.07 &  18.04 &  17.92 &  17.82 &  18.06\\
      &  0.135 &  0.022 &  0.024 &  0.016 &  0.019 &  0.022 &  0.018 &  0.025 &  0.018 &  0.024 &  0.030 &  0.034 &  0.063\\
 Is50233 &  19.47 &  19.65 &  19.30 &  19.16 &  18.92 &  18.93 &  18.91 &  18.66 &  18.63 &  18.73 &  18.62 &  18.39 &  18.60\\
      &  0.151 &  0.078 &  0.065 &  0.070 &  0.060 &  0.074 &  0.066 &  0.065 &  0.075 &  0.069 &  0.093 &  0.110 &  0.190\\
 Is50286 &     &  21.36 &  20.64 &  20.37 &  19.99 &  20.06 &  20.16 &  19.71 &  19.72 &  19.75 &  19.66 &  19.77 &  19.83\\
      &     &  0.296 &  0.199 &  0.144 &  0.108 &  0.117 &  0.137 &  0.139 &  0.140 &  0.155 &  0.168 &  0.249 &  0.261\\
 Id50357 &     &  20.57 &  20.00 &  19.49 &  19.28 &  19.23 &  19.05 &  18.84 &  18.71 &  18.55 &  18.45 &  18.28 &  18.39\\
      &     &  0.063 &  0.033 &  0.025 &  0.031 &  0.023 &  0.016 &  0.037 &  0.024 &  0.036 &  0.032 &  0.044 &  0.070\\
 Id50401 &  20.74 &  20.62 &  20.32 &  20.17 &  19.70 &  19.69 &  19.69 &  19.45 &  19.28 &  19.19 &  19.19 &  18.80 &  18.96\\
      &  0.539 &  0.116 &  0.096 &  0.095 &  0.096 &  0.099 &  0.099 &  0.137 &  0.105 &  0.104 &  0.152 &  0.096 &  0.195\\
 Id50415 &  21.28 &  19.82 &  19.43 &  19.14 &  19.02 &  18.93 &  18.92 &  18.39 &  18.47 &  18.52 &  18.31 &  18.10 &  18.00\\
      &  1.172 &  0.100 &  0.105 &  0.084 &  0.099 &  0.092 &  0.080 &  0.074 &  0.073 &  0.090 &  0.101 &  0.067 &  0.109\\
 Id50552 &  20.05 &  19.91 &  19.55 &  19.47 &  19.05 &  18.98 &  18.83 &  18.71 &  18.48 &  18.47 &  18.39 &  18.10 &  18.00\\
      &  0.243 &  0.127 &  0.125 &  0.116 &  0.089 &  0.092 &  0.106 &  0.113 &  0.081 &  0.094 &  0.102 &  0.095 &  0.093\\
 Is50696 &  18.88 &  18.70 &  18.43 &  18.26 &  17.99 &  18.01 &  17.89 &  17.71 &  17.55 &  17.51 &  17.45 &  17.25 &  17.41\\
      &  0.122 &  0.054 &  0.044 &  0.036 &  0.026 &  0.024 &  0.036 &  0.045 &  0.050 &  0.049 &  0.059 &  0.059 &  0.098\\
 Id50785 &  19.27 &  19.62 &  19.36 &  19.13 &  18.84 &  18.75 &  18.67 &  18.46 &  18.26 &  18.31 &  18.18 &  17.85 &  18.03\\
      &  0.131 &  0.125 &  0.133 &  0.095 &  0.085 &  0.080 &  0.093 &  0.080 &  0.067 &  0.083 &  0.082 &  0.071 &  0.094\\
 Id50826 &  20.48 &  20.61 &  20.45 &  20.21 &  19.79 &  19.89 &  19.59 &  19.44 &  19.15 &  19.12 &  18.81 &  18.71 &  19.01\\
      &  0.512 &  0.233 &  0.213 &  0.191 &  0.145 &  0.183 &  0.125 &  0.154 &  0.116 &  0.124 &  0.115 &  0.114 &  0.268\\
 Is50861 &  19.99 &  19.49 &  19.37 &  19.04 &  18.94 &  18.95 &  19.06 &  18.70 &  18.63 &  18.53 &  18.66 &  18.47 &  18.21\\
      &  0.335 &  0.179 &  0.186 &  0.143 &  0.140 &  0.168 &  0.190 &  0.172 &  0.155 &  0.139 &  0.188 &  0.177 &  0.180\\
 Is50886 &  18.66 &  18.69 &  18.36 &  18.17 &  17.85 &  17.81 &  17.70 &  17.53 &  17.35 &  17.30 &  17.13 &  17.01 &  16.98\\
      &  0.153 &  0.129 &  0.102 &  0.079 &  0.056 &  0.060 &  0.047 &  0.050 &  0.036 &  0.032 &  0.031 &  0.026 &  0.050\\
 Id50960 &  19.47 &  19.14 &  18.85 &  18.63 &  18.35 &  18.33 &  18.22 &  18.09 &  17.98 &  17.89 &  17.75 &  17.59 &  17.61\\
      &  0.144 &  0.031 &  0.033 &  0.027 &  0.026 &  0.029 &  0.025 &  0.035 &  0.031 &  0.034 &  0.033 &  0.041 &  0.064\\
 Is51027 &  20.19 &  19.71 &  19.47 &  19.38 &  19.13 &  19.14 &  19.12 &  19.09 &  19.00 &  18.90 &  18.86 &  18.50 &  18.55\\
      &  0.268 &  0.037 &  0.027 &  0.024 &  0.036 &  0.027 &  0.022 &  0.048 &  0.043 &  0.054 &  0.038 &  0.069 &  0.093\\
 Id70319 &     &  20.95 &  20.66 &  20.54 &  20.26 &  20.12 &  19.95 &  20.06 &  19.73 &  19.58 &  19.51 &  19.25 &  19.36\\
      &     &  0.105 &  0.068 &  0.061 &  0.078 &  0.069 &  0.032 &  0.097 &  0.044 &  0.086 &  0.060 &  0.109 &  0.200\\
 Id70349 &  20.78 &  20.74 &  20.48 &  20.42 &  19.93 &  19.75 &  19.70 &  19.61 &  19.36 &  19.40 &  19.32 &  19.30 &  19.00\\
      &  0.394 &  0.091 &  0.049 &  0.048 &  0.060 &  0.048 &  0.023 &  0.070 &  0.039 &  0.070 &  0.063 &  0.104 &  0.144\\
 Is80172 &  19.93 &  19.71 &  19.31 &  19.10 &  18.74 &  18.79 &  18.79 &  18.50 &  18.42 &  18.48 &  18.50 &  18.24 &  18.22\\
      &  0.206 &  0.033 &  0.021 &  0.017 &  0.022 &  0.015 &  0.010 &  0.025 &  0.024 &  0.033 &  0.026 &  0.045 &  0.109\\
  SBKHP1 &  19.58 &  19.38 &  18.92 &  18.66 &  18.36 &  18.33 &  18.20 &  17.98 &  17.77 &  17.74 &  17.66 &  17.35 &  17.41\\
      &  0.194 &  0.121 &  0.107 &  0.091 &  0.085 &  0.096 &  0.089 &  0.107 &  0.085 &  0.093 &  0.098 &  0.078 &  0.115\\
  SBKHP2 &  19.98 &  19.67 &  19.38 &  19.03 &  18.80 &  18.91 &  18.76 &  18.45 &  18.46 &  18.39 &  18.29 &  18.08 &  18.06\\
      &  0.359 &  0.317 &  0.348 &  0.262 &  0.259 &  0.345 &  0.325 &  0.333 &  0.336 &  0.342 &  0.352 &  0.338 &  0.421\\
  SBKHP3 &  19.09 &  18.99 &  18.65 &  18.43 &  18.09 &  18.15 &  18.05 &  17.82 &  17.61 &  17.61 &  17.49 &  17.22 &  17.26\\
      &  0.202 &  0.221 &  0.216 &  0.192 &  0.170 &  0.226 &  0.197 &  0.226 &  0.179 &  0.201 &  0.197 &  0.167 &  0.226\\
  SBKHP4 &  20.34 &  19.78 &  19.49 &  19.34 &  19.07 &  18.93 &  18.81 &  18.47 &  18.58 &  18.44 &  18.22 &  18.14 &  17.61\\
      &  0.463 &  0.405 &  0.414 &  0.414 &  0.388 &  0.429 &  0.400 &  0.404 &  0.453 &  0.437 &  0.417 &  0.429 &  0.325\\
  SBKHP5 &  19.57 &  19.26 &  18.90 &  18.65 &  18.38 &  18.42 &  18.31 &  18.04 &  17.96 &  17.80 &  17.85 &  17.60 &  17.67\\
      &  0.227 &  0.113 &  0.119 &  0.111 &  0.118 &  0.156 &  0.158 &  0.179 &  0.161 &  0.156 &  0.195 &  0.173 &  0.217\\
  SBKHP6 &  19.08 &  19.08 &  18.76 &  18.58 &  18.25 &  18.22 &  18.15 &  17.89 &  17.89 &  17.71 &  17.74 &  17.47 &  17.56\\
      &  0.169 &  0.254 &  0.267 &  0.262 &  0.252 &  0.292 &  0.294 &  0.334 &  0.331 &  0.310 &  0.360 &  0.336 &  0.483\\
  SBKHP8 &  18.74 &  18.38 &  18.11 &  17.84 &  17.55 &  17.47 &  17.42 &  17.16 &  17.06 &  17.03 &  16.95 &  16.87 &  16.93\\
      &  0.110 &  0.077 &  0.073 &  0.091 &  0.105 &  0.128 &  0.131 &  0.144 &  0.133 &  0.145 &  0.156 &  0.171 &  0.234\\
  SBKHP9 &  19.91 &  19.12 &  18.82 &  18.64 &  18.55 &  18.38 &  18.29 &  18.08 &  18.09 &  18.02 &  17.90 &  17.89 &  18.01\\
      &  0.342 &  0.152 &  0.165 &  0.165 &  0.186 &  0.214 &  0.210 &  0.251 &  0.256 &  0.276 &  0.285 &  0.329 &  0.459\\
 SBKHP11 &  19.25 &  19.56 &  19.21 &  19.08 &  18.82 &  18.76 &  18.69 &  18.55 &  18.35 &  18.28 &  18.44 &  18.38 &  17.94\\
      &  0.144 &  0.179 &  0.182 &  0.179 &  0.190 &  0.217 &  0.207 &  0.274 &  0.234 &  0.239 &  0.337 &  0.366 &  0.308\\
 SBKHP12 &  19.81 &  19.37 &  19.11 &  18.85 &  18.60 &  18.59 &  18.46 &  18.30 &  18.16 &  18.13 &  18.00 &  17.90 &  18.01\\
      &  0.217 &  0.138 &  0.155 &  0.141 &  0.149 &  0.187 &  0.181 &  0.234 &  0.190 &  0.203 &  0.216 &  0.220 &  0.319\\
 SBKHP13 &  19.73 &  19.68 &  19.34 &  19.14 &  18.90 &  18.95 &  18.80 &  18.68 &  18.59 &  18.44 &  18.36 &  18.12 &  18.23\\
      &  0.213 &  0.043 &  0.049 &  0.040 &  0.040 &  0.046 &  0.046 &  0.052 &  0.044 &  0.046 &  0.069 &  0.053 &  0.086\\
 SBKHP14 &  20.08 &  19.52 &  19.40 &  19.09 &  18.91 &  18.79 &  18.84 &  18.64 &  18.53 &  18.62 &  18.40 &  18.35 &  18.60\\
      &  0.336 &  0.194 &  0.207 &  0.173 &  0.174 &  0.189 &  0.210 &  0.241 &  0.204 &  0.257 &  0.242 &  0.265 &  0.415\\
 SBKHP15 &  19.73 &  19.32 &  18.99 &  18.74 &  18.45 &  18.49 &  18.40 &  18.17 &  18.08 &  17.92 &  17.85 &  17.66 &  17.84\\
      &  0.186 &  0.065 &  0.064 &  0.054 &  0.047 &  0.054 &  0.052 &  0.064 &  0.052 &  0.056 &  0.065 &  0.060 &  0.089\\
 SBKHP16 &     &  19.93 &  19.63 &  19.48 &  19.43 &  19.47 &  19.13 &  19.41 &  19.05 &  19.36 &  19.21 &  19.78 &  19.76\\
      &     &  0.179 &  0.186 &  0.106 &  0.195 &  0.248 &  0.128 &  0.333 &  0.154 &  0.363 &  0.343 &  0.712 &  0.838\\
\tableline
\end{tabular}\\
\end{center}
\end{table}

%% file: table2b.tex
\setcounter{table}{1}
\begin{table}
\begin{center}
\caption{Continued} \label{tab correlation}
\begin{tabular}{cccccccccccccc}
\tableline \tableline
$\rm {Cluster}$  & 03  &  04 &  05 &  06 &  07 &  08 &  09 &  10 &  11 &  12 &  13 &  14 &  15\\
(1)    & (2) & (3) & (4) & (5) & (6) & (7) & (8) & (9) & (10) & (11) & (12) & (13) & (14)\\
\tableline
 CFT5 &  20.09 &  20.74 &  20.44 &  20.00 &  19.33 &  19.49 &  19.30 &  19.02 &  18.78 &  18.62 &  18.59 &  18.29 &  18.15\\
      &  0.280 &  0.094 &  0.062 &  0.051 &  0.050 &  0.037 &  0.024 &  0.062 &  0.042 &  0.052 &  0.062 &  0.056 &  0.089\\
    CFT6 &  21.15 &  20.97 &  20.47 &  20.36 &  20.07 &  20.05 &  19.80 &  19.85 &  19.66 &  19.68 &  19.65 &  19.23 &  19.97\\
      &  0.768 &  0.167 &  0.146 &  0.138 &  0.129 &  0.161 &  0.149 &  0.193 &  0.156 &  0.180 &  0.220 &  0.170 &  0.527\\
   CFT37 &  20.79 &  20.32 &  19.69 &  19.71 &  19.51 &  19.05 &  18.94 &  19.10 &  18.67 &  18.93 &  18.38 &  18.15 &  18.79\\
      &  0.644 &  0.334 &  0.266 &  0.277 &  0.274 &  0.207 &  0.181 &  0.313 &  0.191 &  0.329 &  0.142 &  0.179 &  0.501\\
   CFT38 &  20.32 &  19.99 &  19.86 &  19.57 &  19.22 &  19.25 &  19.06 &  18.58 &  18.52 &  18.43 &  18.33 &  18.33 &  18.16\\
      &  0.434 &  0.312 &  0.400 &  0.352 &  0.341 &  0.462 &  0.436 &  0.412 &  0.394 &  0.400 &  0.441 &  0.518 &  0.541\\
   CFT39 &  19.37 &  19.75 &  19.52 &  19.20 &  18.76 &  18.76 &  18.64 &  18.35 &  18.23 &  18.12 &  18.11 &  17.75 &  18.01\\
      &  0.133 &  0.102 &  0.087 &  0.096 &  0.097 &  0.120 &  0.118 &  0.139 &  0.122 &  0.127 &  0.129 &  0.129 &  0.242\\
   CFT41 &     &  20.57 &  19.96 &  19.65 &  19.24 &  19.02 &  18.77 &  19.20 &  19.04 &  18.88 &  18.87 &  19.11 &  19.38\\
      &     &  0.314 &  0.254 &  0.137 &  0.192 &  0.124 &  0.128 &  0.381 &  0.327 &  0.310 &  0.394 &  0.525 &  0.901\\
   CFT97 &  20.95 &  20.16 &  19.97 &  19.70 &  19.64 &  19.62 &  19.41 &  19.30 &  19.48 &  19.50 &  19.56 &  19.15 &  19.14\\
      &  0.542 &  0.200 &  0.181 &  0.149 &  0.159 &  0.207 &  0.182 &  0.205 &  0.221 &  0.256 &  0.316 &  0.252 &  0.290\\
  CFT113 &  22.59 &  20.94 &  20.51 &  20.41 &  20.03 &  20.12 &  19.95 &  19.58 &  19.48 &  19.46 &  19.38 &  19.17 &  19.38\\
      &  2.517 &  0.147 &  0.140 &  0.111 &  0.102 &  0.134 &  0.101 &  0.140 &  0.115 &  0.117 &  0.110 &  0.145 &  0.280\\
\tableline
\end{tabular}\\
\end{center}
\end{table}

%% file: table3.tex
\setcounter{table}{2}
\begin{table}
\begin{center}
\caption{Comparison of Photometry with Previous Measurements}
\label{tab correlation}
\begin{tabular}{ccccc}
\tableline \tableline
Name & $V$ (Previous Work)& $V$ (BATC) &
$B-V$ (Previous Work) & $B-V$ (BATC)\\
\tableline
  Is40083 &  18.39 & 18.373 $\pm$ 0.021 &          0.69 &    0.652 $\pm$  0.043\\
  Is40165 &  18.23 & 18.192 $\pm$ 0.019 &          0.69 &    0.706 $\pm$  0.040\\
  Is40181 &  18.93 & 18.901 $\pm$ 0.030 &          1.09 &    0.970 $\pm$  0.087\\
  Is50225 &  18.43 & 18.439 $\pm$ 0.032 &          0.97 &    0.929 $\pm$  0.065\\
  Is50233 &  19.18 & 19.044 $\pm$ 0.107 &          0.89 &    0.769 $\pm$  0.165\\
  Is50286 &  20.16 & 20.138 $\pm$ 0.194 &          0.89 &  ...            \\
  Id50357 &  19.67 & 19.410 $\pm$ 0.047 &          1.27 &  ...             \\
  Id50401 &  19.93 & 19.908 $\pm$ 0.160 &          1.22 &    0.927 $\pm$  0.302\\
  Id50415 &  19.24 & 19.136 $\pm$ 0.157 &          0.85 &    1.216 $\pm$  0.412\\
  Id50552 &  19.52 & 19.257 $\pm$ 0.157 &          1.14 &    0.885 $\pm$  0.261\\
  Is50696 &  18.13 & 18.120 $\pm$ 0.046 &          0.92 &    0.805 $\pm$  0.101\\
  Id50785 &  19.08 & 19.011 $\pm$ 0.143 &          0.86 &    0.723 $\pm$  0.232\\
  Id50826 &  19.86 & 19.946 $\pm$ 0.268 &          1.08 &    0.797 $\pm$  0.475\\
  Is50861 &  19.69 & 19.015 $\pm$ 0.242 &          0.88 &    0.743 $\pm$  0.381\\
  Is50886 &  18.06 & 18.016 $\pm$ 0.102 &          0.91 &    0.870 $\pm$  0.211\\
  Id50960 &  18.49 & 18.498 $\pm$ 0.044 &          0.86 &    0.905 $\pm$  0.083\\
  Is51027 &  19.36 & 19.255 $\pm$ 0.053 &          0.55 &    0.737 $\pm$  0.115\\
  Id70319 &  20.77 & 20.447 $\pm$ 0.121 &          0.99 &  ...             \\
  Id70349 &  20.12 & 20.198 $\pm$ 0.092 &          0.91 &    0.739 $\pm$  0.210\\
  Is80172 &  18.97 & 18.892 $\pm$ 0.033 &          0.91 &    1.080 $\pm$  0.089\\
   SBKHP1 &  18.54 & 18.516 $\pm$ 0.147 &          1.10 &    1.135 $\pm$  0.238\\
   SBKHP2 &  18.97 & 18.889 $\pm$ 0.459 &          1.02 &    1.045 $\pm$  0.659\\
   SBKHP3 &  18.35 & 18.229 $\pm$ 0.308 &          1.04 &    0.980 $\pm$  0.439\\
   SBKHP4 &  19.24 & 19.253 $\pm$ 0.666 &          1.05 &    0.842 $\pm$  0.895\\
   SBKHP5 &  18.45 & 18.501 $\pm$ 0.206 &          1.04 &    1.037 $\pm$  0.280\\
   SBKHP6 &  18.80 & 18.414 $\pm$ 0.434 &          0.97 &    0.861 $\pm$  0.558\\
   SBKHP8 &  18.01 & 17.713 $\pm$ 0.177 &          1.04 &    0.937 $\pm$  0.212\\
   SBKHP9 &  18.76 & 18.680 $\pm$ 0.311 &          0.98 &    0.810 $\pm$  0.407\\
  SBKHP11 &  18.59 & 18.975 $\pm$ 0.320 &          0.82 &    0.720 $\pm$  0.407\\
  SBKHP12 &  18.70 & 18.734 $\pm$ 0.257 &          1.00 &    0.915 $\pm$  0.338\\
  SBKHP13 &  19.12 & 19.006 $\pm$ 0.068 &          0.87 &    0.889 $\pm$  0.122\\
  SBKHP14 &  19.03 & 19.057 $\pm$ 0.293 &          0.92 &    0.736 $\pm$  0.429\\
  SBKHP15 &  18.73 & 18.580 $\pm$ 0.083 &          0.99 &    1.028 $\pm$  0.146\\
  SBKHP16 &  19.83 & 19.478 $\pm$ 0.312 &          0.75 &    1.145 $\pm$  0.799\\
     CFT5 & 19.826$\pm$0.008 & 19.551 $\pm$ 0.079 &1.184$\pm$0.029 &    1.238 $\pm$  0.187\\
     CFT6 & 20.263$\pm$0.010 & 20.219 $\pm$ 0.227 &0.813$\pm$0.035 &    1.027 $\pm$  0.433\\
    CFT37 & 19.882$\pm$0.009 & 19.772 $\pm$ 0.433 &0.661$\pm$0.013 &    0.913 $\pm$  0.688\\
    CFT38 & 19.835$\pm$0.006 & 19.376 $\pm$ 0.609 &0.808$\pm$0.012 &    0.847 $\pm$  0.785\\
    CFT39 & 19.319$\pm$0.005 & 18.958 $\pm$ 0.168 &0.832$\pm$0.009 &    0.884 $\pm$  0.226\\
    CFT41 & 19.721$\pm$0.006 & 19.500 $\pm$ 0.278 &0.696$\pm$0.009 &  ...             \\
    CFT97 & 19.980$\pm$0.008 & 19.709 $\pm$ 0.276 &0.870$\pm$0.023 &    0.792 $\pm$  0.453\\
   CFT113 & 20.192$\pm$0.014 & 20.173 $\pm$ 0.183 &1.065$\pm$0.058 &    1.355 $\pm$  0.754\\
\tableline
\end{tabular}\\
\end{center}
\end{table}